\documentclass[sigplan,10pt]{acmart}
\renewcommand\footnotetextcopyrightpermission[1]{}
\usepackage{filecontents}
\usepackage{graphicx}
\usepackage{enumitem}
\usepackage{url}

\begin{document}

\settopmatter{printacmref=false}

\date{}

\title{Singularity:  Planet-Scale, Preemptive and Elastic Scheduling of AI Workloads}

\author{Dharma Shukla, Muthian Sivathanu, Srinidhi Viswanatha, Bhargav Gulavani, Rimma Nehme, \\
Amey Agrawal, Chen Chen, Nipun Kwatra, Ramachandran Ramjee, Pankaj Sharma, \\ 
Atul Katiyar, Vipul Modi, Vaibhav Sharma, Abhishek Singh, Shreshth Singhal, \\
Kaustubh Welankar, Lu Xun, Ravi Anupindi, Karthik Elangovan, Hasibur Rahman, Zhou Lin, \\
Rahul Seetharaman, Cheng Xu, Eddie Ailijiang, Suresh Krishnappa, Mark Russinovich \\
{\em Microsoft}}

\newcommand{\eg}{\textit{e.g.}}
\newcommand{\ie}{\textit{i.e.}}
\newcommand{\etal}{\textit{et al.}}
\newcommand{\etc}{\textit{etc.}}
\newcommand{\adhoc}{\textit{ad hoc}}
\newcommand{\msft}{Microsoft}
\newcommand{\aisc}{Singularity}
\newcommand{\dint}{$D_{Int}$}
\newcommand{\saint}{$SA_{Int}$}
\newcommand{\csaint}{\textit {client $SA_{Int}$}}
\newcommand{\ssaint}{\textit {server $SA_{Int}$}}
\newcommand{\tideal}{$T_{ideal}$}
\newcommand{\treal}{$T_{real}$}
\newcommand{\nvidia}{NVIDIA}

\begin{abstract}

Lowering costs by driving high utilization across deep learning workloads is a crucial lever for cloud providers.
We present {\em \aisc}, \msft's globally distributed scheduling service for highly-efficient and reliable execution of deep learning training and inference workloads. At the heart of \aisc\ is a novel, workload-aware scheduler that can transparently preempt and elastically scale deep learning workloads to drive high utilization without impacting their correctness or performance across a global fleet of AI accelerators (e.g., GPUs, FPGAs).

All jobs in \aisc\ are {\em preemptible}, {\em migratable}, and {\em dynamically resizable} ({\em elastic}) by default: a live job can be dynamically and transparently (a) preempted and migrated to a different set of nodes, cluster, data center or a region and resumed {\em exactly} from the point where the execution was preempted, and (b) resized (i.e., elastically scaled up/down) on a varying set of accelerators of a given type. Our mechanisms are {\em transparent} in that they do not require the user to make any changes to their code or require using any custom libraries that may limit flexibility. Additionally, our approach significantly improves the reliability of deep learning workloads. We show that the resulting efficiency and reliability gains with \aisc\ are achieved with negligible impact on the steady-state performance. Finally, our design approach is agnostic of DNN architectures and handles a variety of parallelism strategies (e.g., data/pipeline/model parallelism).

\end{abstract}

\maketitle
\pagestyle{plain}

\section{Introduction}

\begin{figure}
        \begin{center}
                \includegraphics [width=1.0\linewidth]{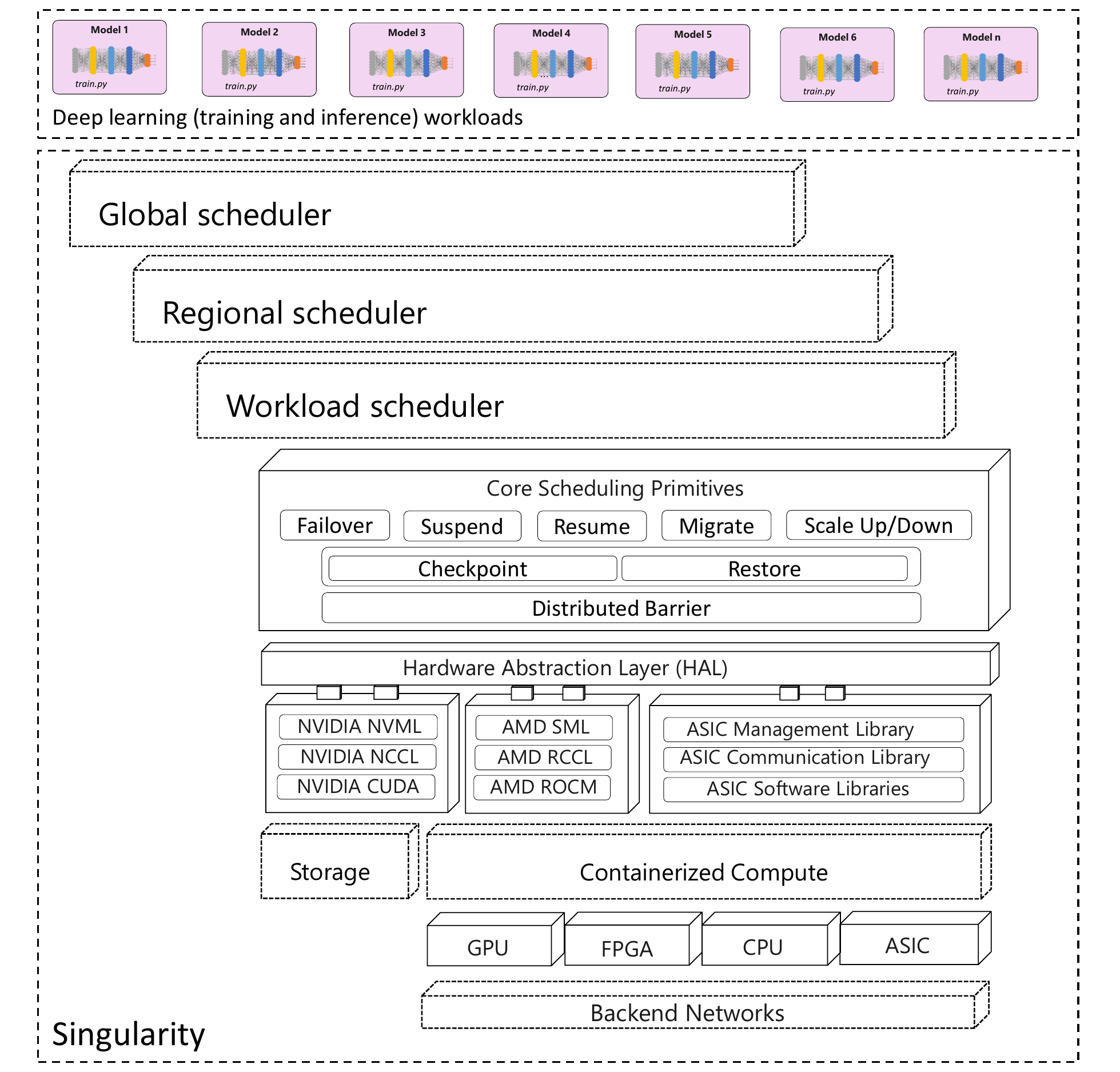}
                \caption{{\bf \aisc\ System Architecture}}
                \label{fig:architecture}
\vspace{-0.2in}
        \end{center}
\end{figure}

\aisc\ is a fully managed, globally distributed infrastructure service for AI workloads at \msft, with support for diverse hardware accelerators.  \aisc\ is designed from the ground up to scale across a global fleet of hundreds of thousands of GPUs and other AI accelerators.  
\aisc\ is built with one key goal: driving down the cost of AI by maximizing the {\em aggregate useful throughput} on a given fixed pool of capacity of accelerators at planet scale, while providing stringent SLAs for multiple pricing tiers.  Figure~\ref{fig:architecture} shows the high-level architecture of \aisc, including its hierarchical scheduling system consisting of scheduling micro-services at the global, regional and workload scopes.

\subsection{Design Goals}

In order to maximize the fleet-wide throughput, \aisc\ adopts the following design principles:

\noindent{\bf a. No idling of resources:}   \aisc\ treats the entire fleet of accelerators as a {\em single logical, shared cluster}, and avoids any resource fragmentation or static reservation of capacity.   The \aisc\ scheduler opportunistically uses spare capacity anywhere across the globe transcending cluster, region, and workload boundaries (training vs. inference).

\noindent{\bf b. Provide job-level SLAs:}  While opportunistically using spare capacity, \aisc\ simultaneously provides {\em isolation} by respecting job-level SLAs.   For example, \aisc\  adapts to increasing load on an inference job, freeing up capacity by elastically scaling down or preempting training jobs.

\noindent{\bf c. Resilience to failures:} DNN training jobs run for hours, days or even weeks, so {\em restarting} from scratch upon failure has a prohibitive cost.   In \aisc, jobs resume from where they got preempted, thus minimizing wasted work.

\subsection{Key Mechanisms}

To achieve the above three goals, \aisc\ relies on two key core scheduling primitives:

\noindent {\bf 1. Preemption and Migration: }\aisc\ can transparently checkpoint, preempt and migrate all DNN jobs across nodes or even across clusters and regions.  The checkpointing is done by using an efficient synchronization barrier to achieve a consistent cut of the distributed state across all the workers of a distributed job.  

\noindent {\bf 2. Resizing/Elasticity: } \aisc\ enables all jobs to be dynamically and elastically scaled up or down in a transparent manner to use a variable number of AI accelerators.

The above mechanisms are {\em transparent} in that \aisc\ requires no changes to the user scripts, or has no dependencies on frameworks/libraries etc. (much like how an operating system preempts and time-slices unmodified processes).   This is crucial for providing consistent SLAs to arbitrary deep learning workloads without relying on any cooperation from users to maintain the SLAs.   As a result, checkpointing, migration, and elasticity are {\em enabled by default for all jobs}, instead of being the niche features they are today.   

The above mechanisms are also {\em work-conserving}, in that a migrated or resized job resumes at exactly the same point in program execution with exactly the same state (e.g., program counter, stack, etc.) as when it was preempted or resized.

While \aisc\ is a significantly broad and complex distributed system, in this paper, we only focus on the above core mechanisms of the \aisc\ scheduler that make all jobs transparently preemptible, migratable, and elastic.

At the core of transparent preemption, migration and elasticity in \aisc, is a novel approach of {\em automatically decoupling} the job from its accelerator resources. The binding between the job workers in \aisc\ and the accelerator devices is dynamic and constantly changing during the lifetime of the job.   To scale up or scale down a job, we simply change the number of devices the workers are mapped to.   This is completely transparent to the user, as the {\em world-size} (\ie total number of workers) of the job remains the same regardless of the number of physical devices running the job.  \aisc\ uses a novel technique called {\em replica splicing} that makes it possible to time-slice multiple workers on the same device with negligible overhead, while enabling each worker to use the entire device memory.  Replica splicing relies on domain knowledge to exploit memory content similarity between workers of a distributed training job at specific points in the program execution.

To achieve such decoupling, \aisc\ introduces the notion of a {\em device proxy}.  The device-proxy runs in its own address space and has a one-to-one correspondence to a physical accelerator device.   When a job worker initiates device APIs, they are intercepted and sent over the shared memory to the device-proxy process,  whose lifetime is decoupled from the lifetime of the worker process.   This separation achieves two key benefits: (a) the host address space is kept clean of device-specific mappings and other side-effects created by GPU libraries such as CUDA, thus making it easier to checkpoint and migrate the host process with an existing tool called CRIU~\cite{CRIU} (b) it allows dynamic, transparent time-slicing of multiple workers on the same device, with the device-proxy performing the multiplexing and scheduling across the workers.

\subsection{Contributions}
In this paper, we make the following key contributions:
\begin{enumerate}
\item

We propose a transparent and robust mechanism to checkpoint generic DNN jobs that are written without checkpointing support, thus {\em making all jobs automatically checkpointable, preemptible and migratable.}

\item

We propose novel, semantics-aware techniques for achieving {\em distributed barrier} across all the workers of a DNN job in a completely transparent manner. This is key to achieving a consistent cut of the distributed state of a DNN training job (including, program counter, stack, CPU and GPU memory etc) spanning multiple machines, so that it can be resumed precisely from the same point when its execution was previously preempted.

\item

We enable, for the first time, transparent and dynamic resizing {\em that makes all jobs elastic by default}, using a novel technique called {\em replica splicing} that allows transparent time-slicing of multiple workers of a DNN job on the same GPU with negligible overhead ( 2-3\%) with ultra-lightweight context switches in a robust and generic manner.

\item

We evaluate the efficacy and the end-to-end efficiency of our core mechanisms, with detailed experiments across models employing different kinds of parallelism (data, pipeline, or tensor parallelism).
We show that the steady state performance overhead of dynamic interception of GPU calls,  the transparent distributed barrier algorithm, and the transparent time-slicing, is negligible (all within 3\%) for a wide range of workloads.  Further, we demonstrate that this approach is robust and practical to maintain despite rapidly changing versions of PyTorch and CUDA. We also show that migration and resizing latency in \aisc\ is reasonable (tens of seconds) with no repeat of computation, and that checkpoint sizes are comparable to user-level checkpointing.

\end{enumerate}

The rest of the paper is structured as follows:  In \S~\ref{sec-overview} we provide an overview of the core scheduling mechanisms.  We present our basic approach of domain-specific interception in \S~\ref{sec-design}, before describing the design of transparent migration (\S~\ref{sec-migration}) and transparent elasticity (\S~\ref{sec-elasticity}).  We discuss the implementation in \S~\ref{sec-implementation}, and evaluate our system in \S~\ref{sec-evaluation}.  We discuss related work in \S~\ref{sec-related}, and conclude in \S~\ref{sec-conclusion}.

\section{Overview of Key Mechanisms}
\label{sec-overview}
In order to improve utilization and reliability, Singularity introduces novel mechanisms to make all jobs preemptible and resizable by default.   \aisc\ takes a consistent checkpoint of a job distributed across several nodes and employing various kinds of parallelism such as data, pipeline, or tensor parallelism, and can resume the job at a later point on a potentially different number of devices using transparent time-slicing, in a potentially different region.   In this section, we first describe three key aspects of these mechanisms:  (a) transparency (no change or constraints on the user code),  (b) work-conserving  (job resumes from the same program execution point where it was previously preempted at), and (c) decoupled execution.  We then describe how the scheduler exploits these mechanisms.

\subsection{Transparent to Users}
\label{subsec-transparency}

Existing approaches for checkpointing and elasticity rely on the user to either directly write code to implement these mechanisms, or to use specific libraries~\cite{deepspeed, pytorch-elastic} that handle checkpointing and elasticity.   Both these approaches are sub-optimal. The former burdens the user with the complexity of saving/restoring python program state (\eg, loop variables, control flow, learning rate scheduler, dataloader state, instruction pointer, \etc), ensuring that the program resumes at the correct point (much like writing re-entrant code, which is error-prone and hard to debug), changing hyper-parameters when the job scales up or down, and so on.  In the latter approach, the user loses flexibility, as these libraries take control of the training loop to keep checkpointing tractable.   This restricts the customizability to the user; such libraries have poor adoption as a result.  Most DNN training workloads today as such are not checkpointable or resizable.

\aisc\ represents a paradigm shift from the status quo, in that our core mechanisms of checkpointing, migration, and elasticity do not require any cooperation from the user, and are automatic, by default.  This is important for two reasons: (1) it enables the scheduler to rely on these mechanisms as first-class constructs for {\em all} jobs in order to provide stringent SLAs and, (2) it completely hides the complexity of checkpointing and elasticity from the user who can focus on writing code with full flexibility, and not even be aware of these mechanisms.  For example, the user writes code for a constant {\em world-size} (\ie, number of workers), and is oblivious to how many physical devices \aisc\ places the job on.

\subsection{Work-conserving}

The checkpoint that \aisc\ takes is comprised of consistent {\em address-space snapshots} of individual workers of the job.  As these snapshots capture the full program state such as instruction pointer, stack, heap etc., the job resumes exactly from the point where it was preempted, with no lost work.  In contrast, today's mechanisms for checkpointing and elasticity force the program to restart from a previous model checkpoint, thus redoing the initialization work and the work performed since the last checkpoint (usually at an epoch-granularity to keep the checkpointing logic tractable).

\subsection{Decoupled Execution}

The goals of improving the reliability of the jobs and the efficiency/utilization of the fleet, are extremely synergistic. The \aisc\ scheduler addresses these goals by {\em decoupling the mapping between the jobs from the underlying resources}. The \aisc\ scheduler transparently virtualizes the {\em world size} and the {\em rank assignment}. This decoupling is crucial for transparently checkpointing and preempting jobs and subsequently resuming them on different nodes, clusters, data centers from the previous checkpointed state, with either the same or different number of GPUs. During the steady state execution of a job, the Singularity scheduler also transparently decouples the training logic of the job worker process (\eg, written in PyTorch) from its interaction with the GPU (Section~\ref{sec-design}).

\subsection{Benefits to Users}

Transparent and work-conserving checkpointing, migration, and elasticity empower the scheduler in fundamental ways:

\noindent{\bf Improved fault tolerance.}  Any job (irrespective of whether the user has written the checkpointing logic) can be resumed precisely from the most recent checkpoint taken automatically by the system in the event of hardware failures (GPU/node/network),  instead of restarting from scratch. This significantly improves the useful fleet-wide throughput.

\noindent{\bf Opportunistic usage of capacity. } 
Transparent preemption and migration enables a job to use spare resources anywhere regardless of cluster/region boundaries. Further, it enables jobs with lower SLA to opportunistically use spare capacity, and be quickly preempted (without lost work) when jobs with higher SLA arrive. Transparent elasticity enables jobs to {\em expand} to use spare capacity and shrink when capacity becomes scarce. 

\noindent{\bf Background defragmentation for locality.}  Locality or topological domains for fault-tolerance (e.g., racks), device-to-device interconnectivity, data locality etc. get fragmented as jobs enter and leave the system, making it hard to schedule a large job with locality constraints. 
Migration of small jobs enables the scheduler to defragment locality domains to place larger jobs. 

\noindent{\bf  Online upgrades.}  Fleet-wide live upgrades can be done without killing jobs, as jobs running on those machines can be cheaply and transparently migrated to a different cluster.

\begin{table}
	\begin{center}
		\footnotesize{
				\begin{tabular}{ l | c |  c | c }
					\hline
					{\bf Metric} & {\bf Premium} & {\bf Standard} & {\bf Basic} \\
					\hline 
					GPU time fraction &	95\%* & 70\%* & Best effort  \\
					guarantee  & & & (like spot VMs) \\
					\hline
                Preemption	& Almost never	& Infrequent	& Frequent \\
                \hline
                Scale-up priority 	& High	& Medium	& Low \\
                (spare capacity) & & & \\
                \hline
                Scale-down priority 	& Low	& Medium	& High \\
                (capacity crunch) & & &  \\
					\hline
					\hline
				\end{tabular}
		}
	\end{center}
	\caption{{\bf SLA for Training Jobs.} Time-fraction values are illustrative; exact values are still evolving.}
	\label{table-sla}
\vspace{-0.3in}
\end{table}

\subsection{Throughput SLAs for training efficiency}

Transparent job preemption, migration, and elasticity enable \aisc\ to define novel SLA for training efficiency. While traditional SLAs such as latency or five-nines of availability apply to inference workloads,  they are a poor fit for DNN training.   \aisc\ introduces the {\em GPU fraction} metric that quantifies throughput in the face of preemption and elasticity.  Table~\ref{table-sla} describes the multiple tiers of SLAs that Singularity provides. A job arriving with a demand for $N$ GPU (based on soft quota) may get more than $N$ or fewer than $N$ GPUs, depending on the competing cluster load. The user is charged only for the actual usage and not for the quota; unused quota can be transparently used by other users/jobs.   

If \tideal\ is the wall clock completion time (for all iterations/epochs) of a job when it ran on $N$ dedicated GPUs with no preemption, and \treal\ is the actual completion time in \aisc, \treal\ >= \tideal\  with resource over-subscription, because the job may be preempted or scaled-down by \aisc\ during execution. The GPU time fraction of a job is calculated as \tideal/\treal.   It’s a {\em relative} slowdown, i.e., if the job would take $H$ hours in a dedicated capacity setup, for Premium SLA job it would take at most $H/0.95$ hours, and for Standard SLA tier at most $H$/0.7 hours.  The GPU fraction SLA is enforced at an hourly granularity. The scheduling policy in Singularity is aimed at maximizing the fleet-wide throughput while minimizing violations of these SLAs.

\section{Domain-specific Interception}
\label{sec-design}

To provide transparent checkpointing and elasticity, \aisc\ exploits the narrow interface that exists between the execution on the CPU and the execution on accelerators, such as GPUs.  Any interaction with an accelerator has to go through specific libraries (\eg, CUDA for \nvidia\ GPUs, ROCm for AMD GPUs, \etc.,), which \aisc\ dynamically intercepts via the LD\_PRELOAD mechanism.   Most of this functionality resides in a component called the {\em device-proxy}.  The device-proxy can be viewed as a hardware abstraction service for the accelerator device, and has a server component (one per device), and a client component that is embedded in each process interacting with the device.  All accelerator-specifc APIs invoked by the host are intercepted and shipped to the device-proxy server, which runs in an isolated address space\footnote{The same principle of remote execution of GPU calls is also used in services such as \nvidia\ MPS~\cite{nvidia-mps}, which intercepts on a closed (internal) API.}.  Running device APIs in a separate address space helps in two ways: (a) it keeps the host address space clean of device mappings and other such dependencies that break checkpointing utilities such as CRIU (b) it allows the device-proxy to be efficiently shared across multiple worker processes during time-slicing for elasticity.  
\begin{figure}
        \begin{center}
                \includegraphics [width=1.0\linewidth]{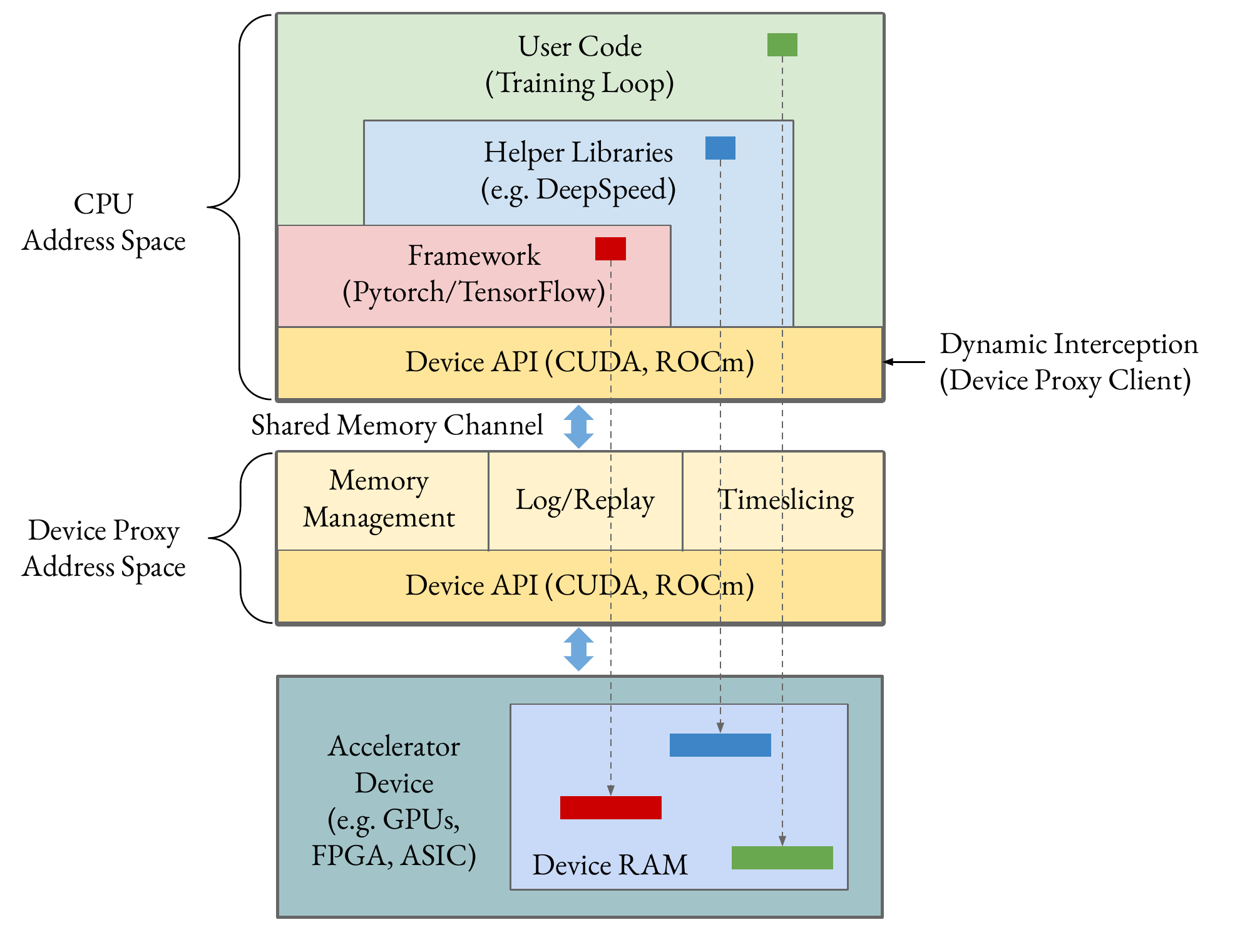}
                \caption{{\bf Overview of Device-Proxy.} Device-proxy runs in a different address space compared to the client process.  CPU address space retains device pointers whose validity needs to be preserved across migration.}
                \label{fig:device-proxy}
\vspace{-0.2in}
        \end{center}
\end{figure}

The communication between the host process and the device-proxy is in the critical path of dispatch to device, so we make it low-latency by using lock-free shared memory channels so that there is no context switch overhead per call.    Figure~\ref{fig:device-proxy} shows the high-level architecture of device-proxy.

There are two types of interceptors in the device-proxy: Dispatch Interceptors (\dint) and Semantics-Aware Interceptors (\saint).  A \dint\ is semantics-oblivious, and only deals with shipping the API cross-address-space to the device-proxy server, handling serialization/deserialization of parameters/response. A \saint, on the other hand, incorporates custom logic at the client-side or server-side (referred to as \csaint\ or\ssaint\ respectively), to implement functionality such as barrier, time-slicing, memory management, etc. Note that \dint\ and \saint\ are not mutually exclusive; for example, the same API may simultaneously have a \csaint, a \dint\ for cross-address-space, and a \ssaint.

\subsection{Limiting surface area of \dint}
 
To be pragmatic and maintainable in a production system, the interception layer must be {\em complete} (all device calls invoked by the job must be intercepted and sent to device-proxy server),  and {\em scalable} (low engineering cost).  Given the rapid pace of new libraries (such as Apex~\cite{Apex}, Thrust~\cite{thrust}, Deepspeed~\cite{deepspeed}, OpenAI Triton~\cite{openai-triton} that directly issue kernels, a naive approach of intercepting all such APIs that interact with the device is impractical.

\aisc\ limits the surface area of \dint's by intercepting at the low-level, \ie. the driver API for launching a kernel (e.g., {\tt cudaLaunchKernel} for \nvidia\ GPUs).  This ensures good coverage while being scalable, as other libraries that define custom kernels eventually go through the launch API.  \aisc\ has an automated code generator that generates the stubs for all \dint's; it simply needs a list of header files from the specific accelerator library (e.g., CUDA), with some annotations to indicate the state-changing calls.

\subsection{Hardware Abstraction Layer for \saint}

While \dint's are auto-generated, most of the custom functionality of the \aisc\ device-proxy (e.g., distributed barrier, context switch for time-slicing) resides in \saint's.  Most of the logic in a \saint\ is device-agnostic, and uses a hardware abstraction layer that maps to device-specific API (e.g., nccl\_allreduce).  The hardware abstraction layer for an accelerator encapsulates key functionality that is common across accelerators.    While the current implementation is specific to \nvidia\ GPUs, to handle a new device type, one simply needs to implement the hardware-abstraction-layer for that device, mapping the specific APIs of the device with the equivalent APIs in the HAL.

There are three types of device-agnostic functionality that require \saint: memory allocation, communication, and device synchronization.  

\subsubsection{Memory allocation}
Memory allocation APIs (e.g., cudaMalloc, cudaFree) need a \saint, because the device-proxy takes over memory allocation.  This enables the device-proxy to have full visibility into what regions of the GPU memory are actually in use, which helps reduce checkpoint size.  It also allows the device-proxy to use custom memory allocation mechanisms to aid transparent time-slicing of multiple workers on the same GPU for elasticity.

\subsubsection{Communication}
Most accelerators have a collective communication library (e.g., NCCL for \nvidia\ GPUs, RCCL for AMD GPUs).  A \saint\ on these APIs enables \aisc\ to implement algorithms for the distributed barrier to synchronize across multiple workers of a distributed job to get a consistent checkpoint (e.g., when no collective calls are in flight at any rank).  \aisc\ provides a generic barrier implementation by piggybacking on the same communication APIs.   \saint\ on these APIs also helps in managing collective calls during time-slicing for elasticity. 

\subsubsection{Device Synchronization}
Device synchronization APIs require a \saint\ to handle transparent elasticity.  The time-slicing in \aisc\  is semantics-aware, as communication across time-sliced ranks has to be correctly handled.  Correct handling of the synchronization APIs (\eg, cudaStreamWaitEvent in CUDA) is essential for correctness and liveness during time-slicing. 

\subsection{\saint\ for Host-specific functionality}

In addition to the hardware abstraction layer for the device, \aisc\ also uses a \saint\ for selective CPU libraries; in particular, the libc I/O libraries (e.g., open, read, write, etc.) are intercepted to track/log updates made by the job to the local file system, so that mutated files can be migrated along with the process checkpoint.   
Unlike a \saint\ for a device library API, a host \saint\ does not have a corresponding \dint, and runs in the host address space.

While domain-specific interception is device-agnostic, for simplicity, the sections below focus on \nvidia\ GPUs.

\section{Transparent Migration}
\label{sec-migration}

In Singularity, preemption, resumption and resizing of a running DNN job involves {\em consistent checkpointing} and {\em restoration} of four broad types of state: (a) program state in CPU (\eg, stack, heap, instruction pointer, etc.), (b) model training state in GPU (\eg, model parameters, optimizer state, etc) (c) control state dealing with interaction between the CPU and GPU (active streams, synchronization events), and (d) inter-GPU and inter-node communication state for different types of parallelism (data/pipeline/tensor-parallel, etc.).  

For scheduled migrations, as well as recovery from unplanned failures, Singularity's transparent checkpointing logic is executed in two modes: (1) {\em On-demand} based on an extenal command when the scheduler decides that the job needs to be preempted, and 
(2) {\em Periodically} based on the interval specified by the user (epoch-level or time-based). 

Achieving transparent checkpointing of generic DNN jobs is challenging for several reasons. First, at the time of checkpointing a given job, Singularity must ensure a {\em consistent cut} of the distributed state spanning multiple hosts and GPUs across multiple machines; all workers of a distributed job must be at a safe and consistent state with respect to collective communication (\eg, allreduce). Second, the in-flight state between CPU and GPU (\eg, active handles, device addresses stored in host memory) must be restored consistently, despite the state management being done by proprietary closed-source libraries such as CUDA. Third, the space overhead of checkpointing must be kept low for large, distributed jobs with hundreds of workers.

\subsection{Checkpointing Program State (CPU)}

There are multiple systems that provide address-space migration~\cite{DMTCP, CRIU}, of which CRIU is the most widely used.  However, a key limitation with CRIU is that it does not handle device mappings by processes using GPU. To use CRIU, the host address space must be isolated from device-specific libraries.
Fortunately, the device-proxy architecture (\S~\ref{sec-design}) provides us that isolation.   The device-proxy server is mostly stateless (with some exceptions addressed below) and hence is not checkpointed; it is simply restarted at the destination.

\subsection{Checkpointing device state}

Model state (e.g., parameters) is checkpointed by the device-proxy process via device-to-host {\em memcpy}.  Because of the memory allocation \saint\ in \aisc, the device-proxy knows which regions of the GPU memory are actually {\em in use}, thus significantly reducing checkpoint size.  

One challenge is that when restoring at the destination, the device memory could get mapped to a different address in the new device-proxy server address space, thus invalidating the pointers in the host process.  To avoid this, the device-proxy hogs the entire GPU memory at startup (with some slack for state tracked by the device libraries), and has a \ssaint\ for {\em mmap} performed by the device allocator (e.g., {\tt cudaMalloc}) to always map to the same CPU address.

\subsubsection{Consistency of device handles in CPU}
\label{subsec-handle-virtualization}
 
Similar to memory pointers, the host address space also retains other handles to the device state.  For example, a {\tt cudaStreamCreate} returns an opaque handle that can be used by the host as a reference in subsequent GPU calls. However, as the device-proxy server is started afresh after migration, the handle would not be valid.  In order to preserve fidelity of these handles across migration, we virtualize these handles.  The device-proxy does not return the actual handle returned by the device, but instead, a virtual handle, and remembers this mapping as part of the client state. After restore and replay, the physical handles may change, but the virtual handle remains stable.  All stateful API calls (e.g., creation of context, stream, event, etc.) are annotated, and the \dint\ for those calls automatically log them for replay upon restore.  We keep the log compact by applying a few domain-specific rules.

\subsection{Communication state}

Most communication between workers of a DNN job happen through collective-communication libraries (e.g., NCCL) that handle both GPU-Direct (e.g., NVLink) and cross-node communication.  As these libraries are proprietary, we cannot handle in-flight communication.  Hence, at the time of checkpointing, we quiesce the job to ensure that there are no collective calls in flight. 

Quiescing cannot be done independently by each worker, because of the semantics of collective communications.  For a collective call (e.g., allreduce) to complete, all participating workers must complete that call.   If a worker checkpoints after say the $n^{th}$ allreduce call has returned (thus freezing that worker), another worker may have already issued the $(n+1)^{th}$ call, which would never complete and thus deadlock.   Thus, before checkpointing, all workers must have issued the same set of collective calls.  \aisc\ uses a novel distributed barrier algorithm that achieves this property in a completely transparent manner.

\subsubsection{Distributed barrier}

One production constraint in designing the distributed barrier algorithm was to 
avoid introducing new failure paths, for example, coordinating through an out-of-band channel (TCP, remote storage, etc.).  The barrier algorithm in Singularity piggybacks on the same communication library that the job uses for collective communication, by introducing additional {\em meta-allreduces} to exchange barrier protocol state.   The algorithm needs to ensure that the ordering of the additional meta allreduces relative to the regular allreduces performed by the job is the same across all workers ({\em program order} requirement for collective communication to avoid deadlocks); 

Our barrier algorithm issues an {\em asynchronous} tandem meta-allreduce before every data allreduce issued by the job.  This trivially ensures consistent program order. The algorithm operates in two phases: {\em Phase 1} is the steady state and {\em Phase 2} is when a barrier request has been received.   The tandem meta-allreduce is a SUM allreduce on a payload that is comprised of two integers:
\begin{itemize}
\item
{\bf need\_barrier:} a worker sends a `1' if it has received a barrier command, `0' otherwise.  If $SUM(need) > 0$, the worker knows someone has initiated the barrier protocol, and switches to {\em Phase 2}.
\item
{\bf ack\_barrier:} a worker sends a `1' if it has switched to {\em Phase 2}, \ie, it acknowledges that it has seen a barrier request directly or indirectly, `0' otherwise.  If $SUM(ack) == world\_size$ (i.e., the total number of ranks), the worker knows that everyone has acknowledged, and can safely acquire the barrier.
\end{itemize}

Once a worker enters {\em Phase 2}, it enters a {\em synchronous mode}: every collective call performed by that worker is synchronous; this ensures timely termination of the barrier protocol.   
The barrier algorithm is guaranteed to complete within at most two mini-batches and guarantees that no in-flight collective calls are outstanding at the time of barrier.  It has very little overhead in the steady state, because the tiny (2-byte) meta-allreduce is asynchronous during {\em Phase 1}.

While the above algorithm works for data-parallel jobs, there is additional complexity with tensor-parallel and pipeline-parallel jobs that may perform multiple allreduces across different groups of nodes, in addition to peer-to-peer calls such as send/recv (for pipelining).   While we can extend the above algorithm to reason about the relative ordering between pipeline and data communication, our current design optimizes for simplicity and checkpoint size, by using the domain knowledge:  we identify the end of a mini-batch, at which point there is no in-flight communication either in the tensor-parallel or pipeline-parallel dimension.  We use the same tandem meta-allreduce protocol as above, but only once at the end of a mini-batch (instead of once per data allreduce) to achieve the barrier.   The tradeoff is that the barrier is delayed until the end of mini-batch (few seconds for large models), but it gives us lower checkpoint sizes compared to acquiring barrier in the middle of a mini-batch.

\subsection{File system state}

Workers in a DNN training job sometimes install local packages, and update other local files.  These need to be preserved after migration to a new node.  Performing a container-wide diff of the file system state (relative to a clean base image) would be too expensive.  The host \saint's on {\tt libc} file system APIs help with this: whenever a local file is opened in writable mode, we append the file name to a log, and during checkpoint we copy over those files.  The data copy to remote storage is deduplicated across workers by using content checksums.

\subsection{Checkpoint/Restore flow}
\label{subsec-checkpoint}

After a barrier is acquired successfully, the job is checkpointed by performing {\tt criu checkpoint} on the individual workers.  The CRIU dumps, along with the GPU state dumps of active tensors, are then moved to a remote storage.  At the new destination, on a {\tt criu restore}, the process starts from the exact state it was checkpointed (\ie, just after the barrier was acquired in the device-proxy client).   The first operation the device-proxy client performs is to respawn a fresh device proxy-server, and then {\em replay} the state-changing calls to bring the GPU to the same state as it was before checkpoint.   The device-proxy server also copies back the GPU tensors to GPU RAM at the same addresses they were before the checkpoint.    The device-proxy finally performs a fresh {\em rendezvous}, so that the ranks can discover each other's new locations and re-establish the communication ring.  
In addition to on-demand checkpointing initiated by the scheduler, each job in \aisc\ takes checkpoints at a user-specified frequency (\eg, every 30 minutes) to handle unplanned failures.

\subsection{Compressing checkpoints}
\label{subsec-checkpoint-size}

\aisc\ adopts several techniques to reduce checkpoint size.  First,
\aisc\ performs per-buffer content checksumming to de-dup across workers.  Before uploading a device buffer, a worker computes its content checksum, and uploads the buffer only if no other worker has uploaded the same buffer.  With this, the size of GPU dumps in \aisc\ is similar to that of the user-level checkpointing.  CRIU dumps of CPU address space are deduped {\em across space and time}.  First, there is high overlap of content between the main training process and data loader processes; we intercept the {\tt write} calls made by CRIU to perform content-hash-based dedup of pages.  Second, there is a high degree of overlap between checkpoints of the same process taken at different points of time (as little of the address space changes); dedup in the temporal dimension makes the subsequent {\em incremental} checkpoints much smaller than the first CRIU checkpoint.

\section{Transparent Elasticity}
\label{sec-elasticity}

\begin{figure}
        \begin{center}
                \includegraphics [width=1.0\linewidth]{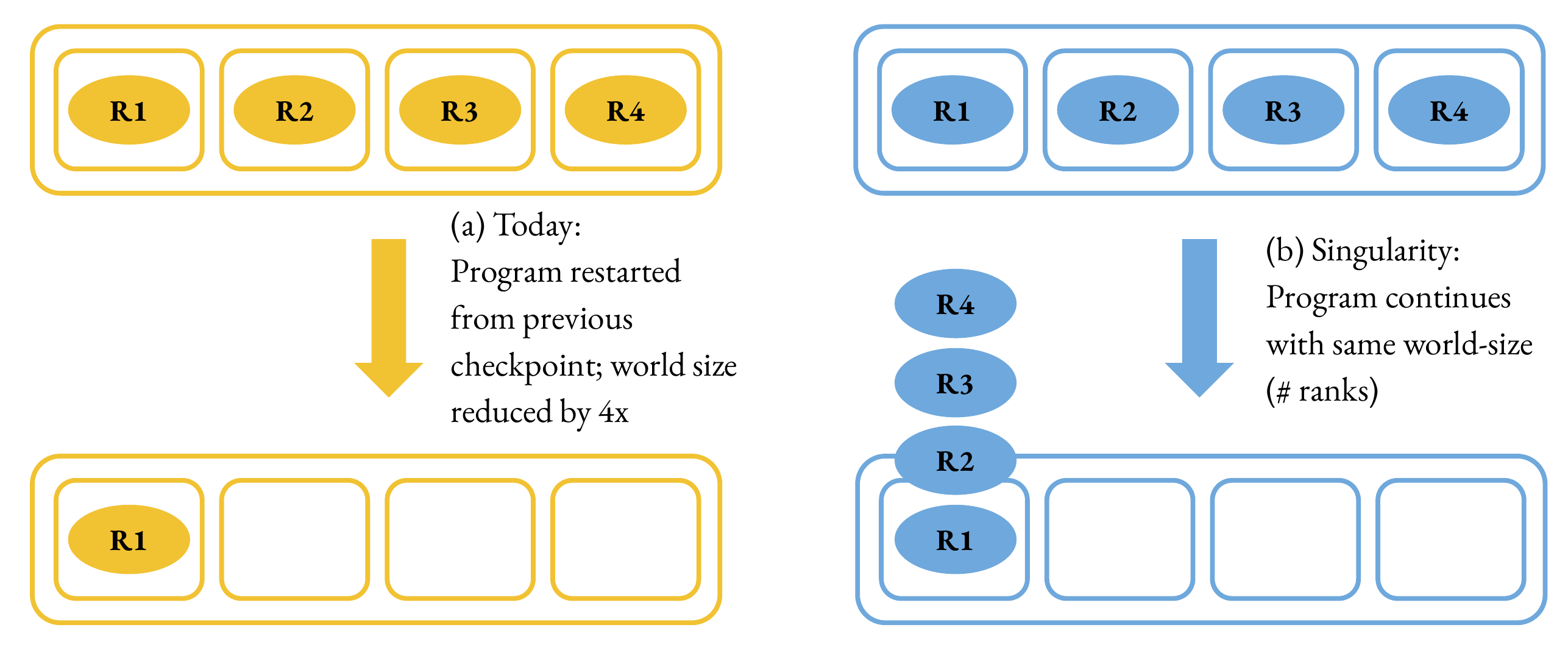}
                \caption{{\bf Elasticity in \aisc\ vs. today}}
                \label{fig:elasticity}
                \vspace{-0.2in}
        \end{center}
\end{figure}

\aisc\ introduces a new ability to resize any DNN training job to use a varying number of GPUs, without requiring any changes to user code and without affecting the semantics of the job.  Resizing a job in Singularity is completely transparent: to the user, the job is always running with the same {\em world size} (\ie, number of ranks/workers).   The scheduler can map each worker {\em one-to-one} to a physical GPU (fully scaled-up), or use a {\em many-to-one} mapping, where a physical GPU is {\em virtualized and time-sliced} across multiple workers (scaled-down case).   In contrast, with libraries such as Deepspeed~\cite{deepspeed} or PyTorch elastic~\cite{pytorch-elastic}, elasticity is exposed to the user, as the job is {\em restarted} with a different {\em world size} after a resize from a previous checkpoint, resulting in wasted work (\eg, initialization and iterations since the previous checkpoint are redone). Figure~\ref{fig:elasticity} illustrates this.

Transparent elasticity builds on top of the transparent migration support in \aisc.   For example, to scale down a job from 4 GPUs to 1 GPU, we simply take a CRIU checkpoint of 3 of the ranks, and migrate those processes to the single GPU with time-slicing.  Because of the properties of CRIU checkpoints, workers resume from exactly the same program state without redoing any computation, so the resize is {\em work-conserving}.

Transparent and dynamic resizing poses several technical challenges. First, when {\em time-slicing} multiple workers of a training job on the same GPU, the fine-grained communication between the workers (\eg, allreduce) must proceed as if the workers were running in different GPUs; this requires the time-slicing to be semantics-aware as well as fine-grained (multiple context switches per mini-batch).   Second, for large models, each worker may utilize nearly the entire RAM on the GPU; fitting multiple workers on the same GPU requires swapping GPU state back-and-forth to host memory, an expensive operation (\eg, 3-10x overhead).   Third, to support transparent elasticity for jobs that use a combination of data-parallelism, pipeline-parallelism and tensor-parallelism, one needs careful placement of workers on GPUs such that only data-parallel replicas of the same model-parallel shard are time-sliced on the same GPU, and to prevent deadlocks in communication scheduling.

\subsection{Semantics-aware time-slicing}

To share a GPU across multiple workers, a simple approach is to run the processes independently on the same GPU.  With this approach, each worker gets its exclusive subset of the GPU memory to store its model state.  However, this approach is a non-starter: in large models, each worker utilizes nearly the entire GPU RAM, so running multiple such workers would simply run out of memory.   Therefore, the time-slicing in \aisc\ needs to be semantics-aware.

The device-proxy  in \aisc\ makes such time-slicing feasible.  Because the device-proxy is decoupled from the host process, we share the same device-proxy across multiple host-processes (\ie, multiple ranks).  As all interactions with the GPU go through the device-proxy, it {\em schedules} the multiple ranks intelligently, allowing only one rank to execute at a given time on the GPU, and then chooses specific points at which to {\em context switch} to another rank.  Conceptually, at the time of context switch, the device-proxy swaps out the GPU memory used by the original rank (\ie, copying to host memory) and then swaps in the GPU memory for the new rank, thus enabling each worker to use nearly the entire GPU RAM.  Of course, such swap-out/swap-in would be quite expensive; we describe how we optimize this in \S~\ref{subsec-splicing}.

To keep overheads low, we must context-switch only when absolutely necessary. When a rank is performing just compute on its respective data (\eg, forward and backward pass operations such as matrix multiplications), there is no need to context-switch.  After the backward pass, the data-parallel ranks participate in collective communication (\eg, allreduce) to exchange gradients, which requires all ranks to participate (and contribute their respective gradients), necessitating a context-switch.   Note that within a single mini-batch, multiple asynchronous allreduce calls may be issued by the framework (to overlap compute with communication).    After allreduces for the mini-batch have been issued, frameworks such as PyTorch typically perform a synchronization operation ({\tt cudaStreamWaitEvent}) on the GPU before copying back the averaged gradients.   At this synchronization point, the device-proxy switches to the next rank that is sharing the GPU and lets it run exclusively until it hits the same point, and then context switches to the next rank, and so on.

Collective communication (\eg, allreduce) happens through proprietary libraries such as NCCL.  NCCL has the notion of a {\em communicator} which is initialized for a particular ring of participating ranks, and subsequent operations such as {\em allreduce} just reference the communicator.  To keep the interaction of NCCL communicators with user-level time-slicing manageable, we decouple the logical data-parallel world size of the job from the world-size that NCCL sees; in our approach, NCCL sees only one rank per GPU. During time-slicing, the device-proxy transparently performs {\em local accumulation} into its scratch buffers, and only the last rank sharing the GPU performs the actual {\tt nccl\_allreduce} with the result of the locally accumulated gradients.    Thus, after a resize operation, the world size seen by NCCL is changed, (handled by the fresh rendezvous after restore (\S~\ref{subsec-checkpoint}).

\subsection{{\em Replica splicing} for memory sharing}
\label{subsec-splicing}

While the above approach for time-slicing is sufficient from a correctness perspective, it is prohibitively slow.  A V100 GPU has 32GB of RAM (A100 is even higher - up to 80GB), so a large model could use most of the RAM for a single worker. A single context-switch (swap-out followed by swap-in of 32GB each from host memory) per mini-batch would take 2-4 seconds, while the mini-batch computation time itself could be much less (hundreds of ms), \ie, a 5-10x overhead.
We now present a new technique called replica splicing that makes context-switches much cheaper.

\subsubsection{Checksum-based dynamic dedup}

The GPU memory consumed by a training job falls into four categories: 
\begin{enumerate}
\item

\noindent{\bf Parameters ({\em P}).}  The weights/parameters for each layer of the model; forward and backward pass run on these tensors.

\item
\noindent{\bf Optimizer State ({\em O}).} State tracked by the optimizer to compute the delta to apply on parameters each iteration.  Tracks historic state (\eg, first and second moments of gradients)

\item
\noindent{\bf Gradients ({\em G}).} Each replica has its own copy of gradients corresponding to its mini-batch.  After the backward pass, the gradients of all replicas are averaged, which is then used to update weights consistently.

\item
\noindent{\bf Activations ({\em A}).}  The intermediate output of the forward pass for each layer; 
used during backward pass to compute gradients relative to input for back-propagation.

\end{enumerate}

The key insight that replica splicing exploits is that among data-parallel replicas, the parameters ({\em P}) and optimizer state ({\em O}) proceed in lock-step, \ie, they are updated at the end of the mini-batch consistently by all replicas, with the (same) averaged gradients.  Thus, at the end of a mini-batch, the tensors corresponding to {\em P} and {\em O} will be identical across ranks sharing the GPU. Further, at the end of the mini-batch, activation tensors ({\em A}) are freed by the framework, as the backward pass is completed.  We use these insights to make the swap-out/swap-in {\em conditional}, in the following manner.

As the device-proxy controls the memory allocator, it has visibility into each buffer that is allocated by the framework.  During a context-switch, the device-proxy computes content checksums for each live buffer.   During swap-out, it first looks up whether the host already contains a buffer with the same content checksum; if yes, it avoids the swap-out and simply marks the GPU buffer as unused (GC happens lazily when there is demand for fresh allocations by the new rank, so that we can opportunistically cache multiple versions in the device).  Similarly, during swap-in of a new rank's data, it checks whether the device already has a buffer with that checksum; if so, it avoids the swap-in from host.  Note that although the content matches, that buffer may be mapped to a different device address in the new rank; in that case, the device-proxy performs a {\em device-to-device move} of that buffer to the desired address which is cheaper than a swap-in from host (HBM bandwidth vs. host memory bandwidth).

With the above optimization, if 4 ranks are sharing a single GPU, the {\em swap-out} of {\em P} and {\em O} buffers during context-switching needs to be done only for the first rank; others would find that the checksums are already present in host memory and omit the swap.  However, note that the swap-in must still be done for each rank; when a rank starts its time-slice, its local state contains {\em P} and {\em O} from the previous mini-batch, while the previous rank's copy is updated to the current mini-batch.  One can avoid the swap-in, if we have space to store two additional versions of {\em P} and {\em O} within the GPU (for any time-slicing factor >2)\footnote {There are at most two versions of {\em P} and {\em O} that are active - current mini-batch and previous mini-batch, and the third copy is needed as scratch space so that the current rank does not overwrite the previous mini-batch's pristine version of {\em P} and {\em O} which would be needed by the next rank.}  This brings up two challenges: (a) the additional space for two extra copies of {\em P} and {\em O} is unacceptable for large models (b) we would still need to perform device-to-device copies of {\em P} and {\em O} during context-switch, as each rank may have allocated  the same buffers at different addresses.  The D2D copy cost is still non-trivial because cyclic dependencies between (thousands of) source and destination buffers, force the copy to happen in phases, thus limiting parallelism.  We next describe how we address these two challenges.

\subsubsection{Using domain knowledge for consistent allocations}
\label{subsec-consistent-alloc}

We first address the problem of the same {\em P} and {\em O} buffers getting different addresses in different ranks; this happens because each worker performs its own allocations.  Singularity uses domain knowledge about deep learning training to make the addresses consistent, without requiring explicit coordination between replicas.  We use the insight that within data-parallel replicas, the allocation sequence (size, ordering) for ``stable'' buffers such as {\em P} and {\em O} (which are preserved across mini-batches) must be the same across all replicas by definition because they have identical sets of parameters.   However, there could be other allocations that are variable-sized across replicas (\eg, activations whose size depends on input data size which may vary across mini-batches). Because of such variable-sized allocations, the state of the memory allocator diverges across replicas, causing even the stable buffer allocations (that are interleaved with other allocations) to get misaligned addresses.

To handle this, the device-proxy in Singularity uses a bi-directional memory allocator.  Stable buffers (such as {\em P} and {\em O}) get allocated at the high end of the address space, while other buffers get allocated at the low end.  This ensures that instability in the transient allocations (such as activations) does not affect the memory allocator metadata in the high region, thus ensuring that the stable buffers such as {\em P} and {\em O} get the same addresses across replicas.  We have empirically validated this across a wide range of models and PyTorch versions.  To identify stable buffers such as {\em P} and {\em O}, we seed the allocator with a list of pre-identified stack traces (Python and C++) pertaining to parameter and optimizer state allocations; this list needs to be updated once per version of PyTorch, and is quite straightforward.  At the time of allocation, the device-proxy client gets the stack trace and matches against this list.  Note that in pathological cases, if the annotations are incorrect and we are not able to get consistent addresses for {\em P} and {\em O}, it can only affect performance (measurable), but not correctness.

\subsubsection{Squashing selective operations}
\label{subsec-squashing}

To avoid dealing with multiple copies of {\em P} and {\em O} (which causes either swap-in cost or extra GPU memory), we use another domain-specific insight.  By definition, all data-parallel replicas will arrive at the same version of {\em P} and {\em O} buffers after the completion of a mini-batch.  We also know that {\em P} and {\em O} buffers are updated only after the allreduce of gradients across replicas is completed.  Thus, if we can identify the operations that update parameters and optimizer state, we can perform those operations only in one of the ranks sharing the device (the ``root'' rank), and simply ``squash'' those operations in other ranks, because (a) they would anyway result in the same final state, and (b) the buffers have the same corresponding addresses across ranks (\S~\ref{subsec-consistent-alloc}) so subsequent mini-batch computation will see the right data.   To squash an operation, the device-proxy simply omits issuing {\tt cudaLaunchKernel} to the GPU for those operations.   With such squashing, we avoid swapping in the previous version of {\em P} and {\em O}, as they are no longer updated by other ranks.

Note that with squashing, we exploit domain-specific understanding to alter the execution sequence, so we need to guarantee that it does not affect correctness.  While reasonable models conform to our assumptions in squashing, we should prevent a pathological model from encountering silent corruption/incorrect execution because of violation of these assumptions.   To ensure robustness, we follow an approach of {\em conservative validation}; if validation succeeds, squashing is guaranteed to be safe and correct, but if validation fails, we conservatively disable squashing.  Thus, violation of our assumptions can only affect performance (in which case we disable time-slicing for that model), but never correctness.

For conservative validation, we always run the first mini-batch (and periodically, every $k^{th}$ mini-batch) with squashing disabled (thus incurring the swap-in/out cost); this is guaranteed to be correct execution.  In the validation mini-batch, we enforce/assert our assumptions on the mutation behavior of operations.   At the device-proxy, identifying buffers mutated by an operation is challenging because, for a GPU operation such as {\tt cudaLaunchKernel}, the parameters could be indirect-addressed via multiple levels of GPU buffers (opaque to the device-proxy).   Therefore, validation relies on a novel approach of {\em inferring the effect} of an operation post-facto, using buffer content checksums.   

During the validation mini-batch, we verify that the model conforms to the following invariants:
\begin{enumerate}

\item
All buffer mutations during the {\em squashing window} must be identical across all ranks sharing the GPU.  The difference between the buffer checksum state between two points indicates mutations performed during that interval.  These mutations in the squashing window must be identical in all respects (same addresses, same checksums, same size).

\item
Device-to-host copies performed during the squashing window must copy exactly the same data across all ranks sharing the GPU.

\end{enumerate}

For simplicity, in PyTorch, we use the optimizer step as our squashing window; however, the above validation mechanism can be used to dynamically infer the squashing window (\ie, sequence of operations where the invariants hold).

If the above validation fails, we treat the model as unsafe for squashing and fall back to the swap-based mechanism (if necessary, ``rolling-back'' the job to the last checkpoint where validation suceeded).  Note that if the model has slack space in the GPU RAM, multiple copies of {\em P} and {\em O} could fit, and the model could still run efficiently, but in general, there will be a performance hit.  We thus convert a potential correctness problem into a measurable performance problem.  Of course, from the perspective of the scheduler, a high-overhead time-slicing would be counter-productive for cluster efficiency.  Therefore, we monitor the overhead due to time-slicing, and if it exceeds a threshold (\eg, >5\% of mini-batch time), we disable time-slicing for that model.  This would be a rare scenario but still needs to be gracefully handled for robustness.

\subsection{Handling model-parallel jobs}

The discussion so far has focused on data-parallel jobs.  Handling model-parallel jobs such as tensor-parallel and pipeline-parallel jobs brings up new challenges.  For example, a tensor-parallel job performs an allreduce for every matrix multiplication in forward and backward pass.  If we context-switch for such allreduce, replica splicing would not work because activation tensors would still be live.  Similarly, pipeline-parallel jobs perform peer-to-peer send and receive of activations and gradients across GPUs/nodes for every micro-batch; time-slicing during the micro-batch would cause excessive swaps because of live gradients and activations.

To address these challenges, \aisc\ uses two key techniques: {\em splicing-aware placement} and {\em inferring the intent of collective calls}.   With splicing-aware placement, we ensure that only data-parallel replicas of the {\em same} model-parallel partition are time-sliced on the same GPU.    For example, to run a 8-rank job with 4-way pipelining and 2-way data-parallelism on 4 physical GPUs, \aisc\ would place the two data-parallel replicas of the same pipeline stage in each GPU.  The same applies to 3D parallel jobs; the ranks that are time-sliced within the same device would belong to the same pipeline stage {\em and} the same tensor-parallel partition.  Note that this requires \aisc\ to be aware of the rank assignment logic.  Two popular libraries \nvidia-Megatron~\cite{megatron}, and DeepSpeed~\cite{deepspeed} have the same rank assignment logic across the parallelism dimensions, and this logic is mirrored in \aisc. For jobs that use a custom launcher with different rank assignment policy, \aisc\ provides an API for the job to communicate the rank-to-topology mapping for all ranks (\eg, Rank 4 is DP0, MP0, PP1 \etc).

Second, the device-proxy infers the {\em intent} of collective communications and triggers time-slicing only on collective calls in the data-parallel dimension.  Other collective calls simply pass through without context-switching, which is correct because completion of those calls only depends on ranks executing on other GPUs and does not need input from other data-parallel replicas time-sliced in the same GPU.  However, inferring the intent of a particular allreduce call transparently is non-trivial, as each user model could have its own control flow and ordering across communication in multiple dimensions of parallelism.  \aisc\ leverages the initialization path of collective communication (\eg, {\tt ncclCommInitRank}) to achieve this.  It forces a context-switch after every {\tt ncclCommInitRank}, and the device-proxy (that is shared across all ranks using the same device) keeps a per-communicator count.  After a full round of context switches, if the local count for a communicator is >1, the device-proxy infers that the communicator is in the data-parallel dimension (because of splicing-aware placement above).  During a collective call, it simply looks up a map on the communicator to know whether it is data-parallel.

\subsection{Handling ZeRO-redundancy optimizer}
ZeRO~\cite{deepspeed-zero} shards the data-parallel state such that there is no redundancy across data-parallel workers.  Such partitioning violates our invariants for squashing validation (\S~\ref{subsec-squashing}).  To handle this, \aisc\ introduces the notion of {\em partial sharding} for ZERO, which decouples the sharding factor (minimum needed to fit the model in GPU) from data-parallelism degree (for parallelism).  If the two are equal, the model is by definition not shrinkable to fewer GPUs as it cannot fit.  If the data-parallelism factor is higher, say, 4x the sharding factor), then we can support up to 4-way time-slicing/scale-down.  In this scenario, the partial sharding factor simply becomes another dimension of model-parallelism, and only replicas of the same ZERO-shard are time-sliced.  Introducing partial sharding in DeepSpeed was quite straightforward (around 30 lines of Python code).

\section{Implementation}
\label{sec-implementation}

Our mechanisms for transparent checkpointing, preemption, resumption, and elasticity are an integral part of the \aisc\ scheduler. In this section, we briefly highlight a few implementation challenges.

\noindent{\bf Serializing opaque parameters.}
In our interception-based device-proxy, the \dint\ for {\tt cudaLaunchKernel} is challenging because its signature is opaque, making serialization difficult (the signature is generated internally by \nvidia's nvcc and is invisible to interceptor). To handle this, we have a custom {\em server} \saint\ for {\tt cudaLaunchKernel} which uses {\tt cuObjDump}, a binary utility in the CUDA toolkit that parses the generated kernel library and extracts parameter information.  To avoid the high cost, we cache this information and run {\tt cuObjDump} only on cache misses.   For JIT kernels, we intercept {\tt nvrtcCompileProgram} and extract the param signatures by parsing the PTX that is generated. 

\noindent{\bf Hiding dispatch latency.}
The cross-address space invocation in device-proxy happens in the critical path of operations such as {\tt cudaLaunchKernel} and {\tt cudaGetLastError}, which affects performance.  We use domain-specific optimizations for the most frequent calls.  For {\tt cudaGetLastError}, we opportunistically issue it at the server during every kernel launch and piggyback it along with its response, so that the device-proxy client can return it from the cache when PyTorch issues it.  For {\tt cudaLaunchKernel}, we perform {\em delayed error notification}, the call returns at the client without waiting for response from device-proxy server; the response is read lazily before issuing the next call to the server, thus allowing overlap between PyTorch processing at client and the latency incurred by the server; because PyTorch (and other frameworks) crash when such (rare) errors are encountered, this does not affect the job semantics.

\noindent{\bf Hiding context-switching overhead.}
Switching from one rank to another during time-slicing involves computing checksums of all live device tensors  and comparing them with the other rank's copy, performing buffer moves if necessary (a few ms of CPU activity).  Further, the switching logic depends on the output of the checksum computation which in turn waits for all prior GPU operations to complete, implicitly forcing a device sync.  To avoid incurring this cost in the critical path, \aisc\ performs {\em eager dispatch} of the next rank.   The device-proxy starts servicing the next rank in parallel with the switching logic, thus overlapping useful work by the next rank (CPU logic, dispatch of device operations) with the switching latency.  By careful use of asynchronous ordering primitives such as {\tt cudaStreamWaitEvent}, we ensure that operations for the new rank execute on the GPU only after the switching is complete.  

\section{Evaluation}
\label{sec-evaluation}

We now evaluate the mechanisms for transparent preemption and elasticity in \aisc, across a wide variety of popular models.  We explore the following issues:
\begin{itemize}
\item
Steady-state overhead of device-proxy
\item
Checkpoint size with transparent checkpointing
\item
Overhead of transparent time-slicing for elasticity
\item
End-to-end latency of migration and resizing
\item
Robustness of the transparent fungibility approach
\end{itemize}

\begin{table}
	\begin{center}
		\footnotesize{
				\begin{tabular}{ l | c |  c | c |c }
					\hline
					Model & Library & \#Params & Domain & Parallelism \\
					\hline
					BERT-MRPC~\cite{bert-code} & Huggingface & 109M & NLP & DP \\
					DenseNet169~\cite{imagenet-code} & PyTorch Ex. & 14M & CV & DP \\
					PyramidNet~\cite{pyramidnet-code} & End-user & 24M & CV & DP \\
					GPT-2~\cite{gpt2-code} & Megatron & 1.8 B & NLG & 3D \\
					ResNet50~\cite{imagenet-code} &  PyTorch Ex. & 26M & CV & DP \\
					InternalQ & Internal & 355M  & NLU & DP \\
					InternalT & Internal & 8 B  & NLG & 3D \\
					\hline
				\end{tabular}
		}
	\end{center}
	\caption{{\bf Models used in evaluation.} Legend: DP=data-parallel 3D=Data+Tensor+Pipeline parallel. InternalQ and InternalT are internal production models at \msft.}
	\label{table-models}
\end{table}

The models we use in our evaluation are listed in Table~\ref{table-models}.  All experiments are run on \nvidia\ DGX-2 servers (8 V100 GPUs per node with NVLink connectivity); servers are connected via InfiniBand. Each server is a Xeon Platinum 8168 with 2 sockets of 20 cores each and 692 GB of RAM.  Per-rank mini-batch size is kept constant independent of data-parallelism degree. 

\subsection{Overhead of device-proxy}

\begin{table}
	\begin{center}
				\begin{tabular}{ l | c |  c | c | c| c |c |}
					\hline
					 & \multicolumn{6}{c|}{Time per minibatch (s)}  \\
					{\bf Model} & \multicolumn{3}{c|}{\bf{1-GPU}} & \multicolumn{3}{c|}{\bf{16-GPU}} \\
					& B & DP & \% & B & DP & \% \\
					\hline
					BERT & 0.43 & 0.43 & 0 & 0.42 & 0.42 & 0 \\
					DenseNet169 & 0.26 & 0.26 & 0 & 0.29 & 0.3 & 2.9 \\
					PyramidNet & 0.25 & 0.24 & -3.3 & 0.27 & 0.26 & -4.2 \\
					ResNet50 & 0.20 & 0.20 & 0 & 0.21 & 0.22 & 2.1 \\
					InternalQ & 0.56 & 0.56 & 0 & 0.72  & 0.72 & 0 \\
					\hline
                \bf{Model} & \multicolumn{3}{c|}{\bf{1-GPU}} & \multicolumn{3}{c|}{\bf{32-GPU}} \\
					& B & DP & \% & B & DP & \% \\
					\hline
					GPT-2 & NA & NA & NA & 1.86 & 1.77 & -5.0 \\					
					InternalT & NA & NA & NA & 7.12 & 6.96 & -2.3 \\
					\hline
				\end{tabular}
	\end{center}
	\caption{{\bf Steady-state overhead of device-proxy.} Mini-batch times for two configurations of job size. B = baseline, DP = \aisc\ (Device Proxy). \% overhead is also shown}
	\label{table-proxy-overhead}
\end{table}

We first evaluate the cost of dynamic interception with our device-proxy architecture, where GPU calls are dispatched to a different address space.   Table~\ref{table-proxy-overhead} shows the average time taken per mini-batch for each of our models, with and without device-proxy.  As can be seen, the device-proxy has negligible impact on end-to-end performance, with overheads below 3\% in most models, including GPT-2 and InternalT which use a combination of data-parallelism, tensor-parallelism, and pipeline-parallelism.  Note that for some models, our performance is actually {\em better} because of our optimizations such as overlapping {\tt cudaLaunchKernel} (\S~\ref{sec-implementation}).

\subsection{Checkpoint Size}

\begin{table}
	\begin{center}
		\footnotesize{
				\begin{tabular}{ l | c| c |  c | c | c| c |c |}
					\hline
					 & User- & \multicolumn{6}{c|}{\aisc\ checkpoint (GB)}  \\
					Model & chkpt & \multicolumn{3}{c|}{4-GPU} & \multicolumn{3}{c|}{8-GPU} \\
					&  (GB) & $S_G$ & $S_{Cr}$ & $S_{Cr}^i$ & $S_G$ & $S_{Cr}$ & $S_{Cr}^i$  \\
					\hline
					BERT & 1.3 & 1.26 & 2.09 & 0.027 & 1.27 & 4.19 & 0.052 \\
					DenseNet169 & 0.11 & 0.26 & 4.39 & 2.88 & 0.42 & 8.73 & 5.63 \\
					PyramidNet & 0.19 & 0.19 & 1.05 & 0.029 & 0.19 & 2.09 & 0.057  \\
					GPT-2* & 24.2  & NA & NA  & NA  & 33.05 & 4.8 & 5.2 \\
					ResNet50 & 0.2 & 0.35 & 4.0 & 2.57 & 0.51 & 8.01 & 4.94 \\
					InternalQ & 4.0 & 5.4 & 1.53 & 0.035 & 5.4 & 3.05 & 0.11 \\
					InternalT* & 274 & NA & NA & NA  & 162 & 13.1 & 0.19 \\
					\hline
				\end{tabular}
		}
	\end{center}
	\caption{{\bf Checkpoint Size.} Compares checkpoint sizes of \aisc\ with user-level checkpoints.  $S_G$ is GPU state, $S_{Cr}$ is the size of first CRIU dump (all workers), $S_{Cr}^i$ is size of subsequent (incremental) dumps.  * on GPT-2 and InternalT indicates 32-worker configs (larger model size).}
	\label{table-checkpoint-size}
	\vspace{-0.3in}
\end{table}

We now evaluate the checkpoint size (\S~\ref{subsec-checkpoint-size}) under \aisc, and compare it with user-level checkpointing, under two configurations: 4 workers, and 8 workers.   In general, the size of our checkpoint for an N-worker job is:

$S_G + S_{Cr}$, where $S_{Cr} = N * S_{pwCr}$

\noindent $S_G$ is the size of GPU state (Parameters and Optimizer State) for a {\em single data-parallel replica}, and $S_{pwCr}$ is the {\em per-worker} CRIU dump size.  While the CRIU dump size scales with the number of workers, the GPU size is independent of the number of data-parallel replicas.   Table~\ref{table-checkpoint-size} shows these results.  We can see that $S_G$, the GPU state saved by \aisc\ checkpoint is quite comparable to user-level checkpoints, despite semantics-oblivious buffer-level checkpointing.  For CRIU checkpoints of CPU state, we show two numbers: the size of the first checkpoint, and that of subsequent checkpoints (representative of continuous checkpointing scenario).  Because of the temporal deduplication that \aisc\ performs, the subsequent checkpoints are an order of magnitude smaller than the first checkpoint for many models.  Even for the first checkpoint, the per-worker CRIU dump size ($S_{pwCr}$) is < 1GB in most cases, which is quite manageable even for large jobs.  As expected, the CRIU dump size for 8-workers is twice as large as the 4-worker config.  Note that the numbers for GPT-2 and InternalT reflect 32-worker configurations.  While cross-worker deduplication of CRIU images could be explored to reduce the size further, we find the present numbers are quite acceptable in practice.

\subsection{Replica splicing for Elasticity}

\begin{figure}
        \begin{center}
                \includegraphics [width=1.0\linewidth]{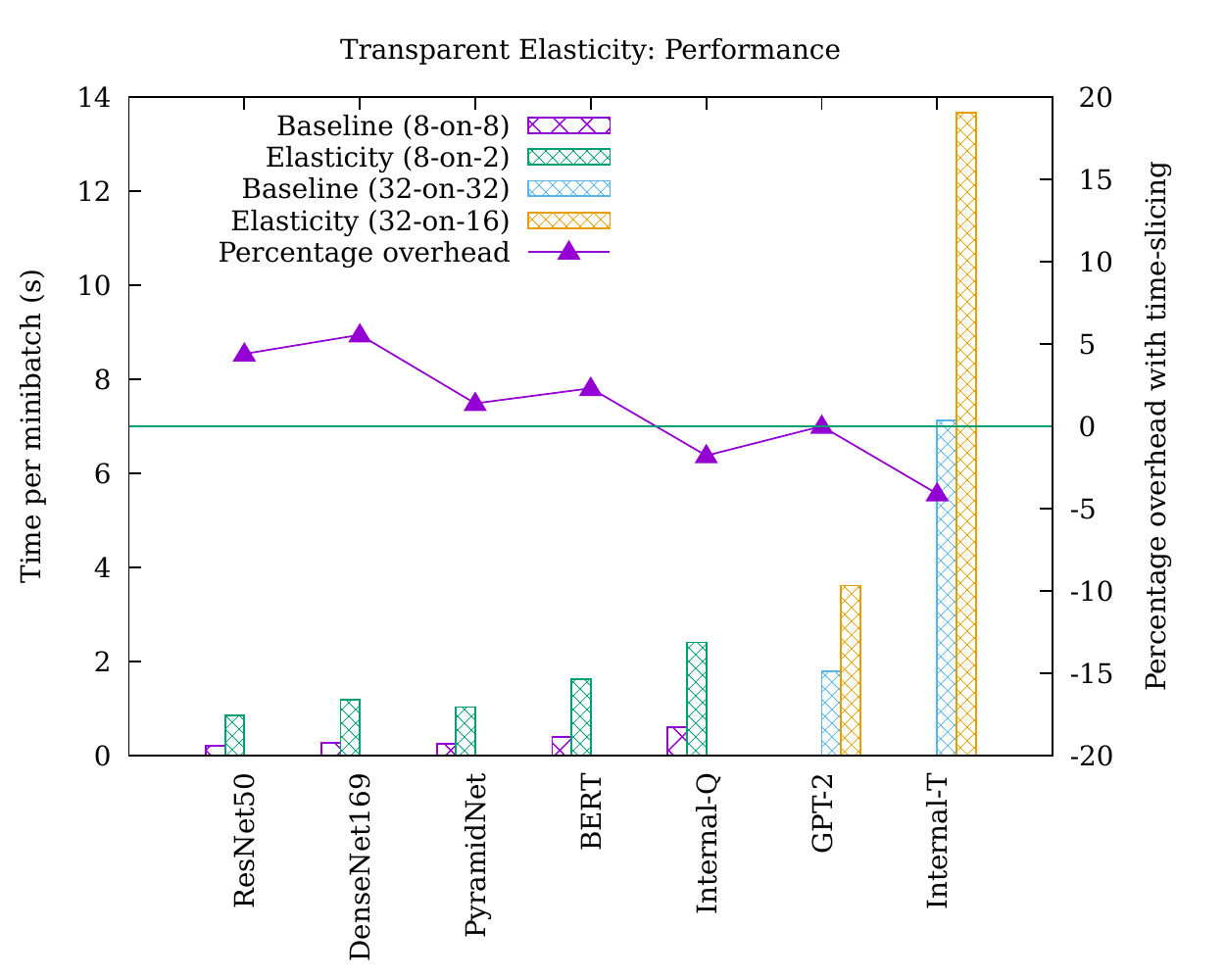}
                \caption{{\bf Overhead of time-slicing}}
                \label{fig:time-slicing-overhead}
\vspace{-0.2in}
        \end{center}
\end{figure}

The core enabler for transparent elasticity is our ability to time-slice multiple workers on the same GPU with replica splicing.  Figure~\ref{fig:time-slicing-overhead} shows the overhead caused by replica splicing for various models, under two configurations: 2-way time-slicing (32 workers on 16 GPUs), and 4-way time-slicing (8 workers on 2 GPUs).  In each, we compare our time-sliced run with a baseline run of unmodified PyTorch (without \aisc) in fully scaled-up mode.  With N-way time-slicing, the mini-batch time is expected to increase by a factor of N (\ie, same work but fewer GPUs); any increase beyond this is overhead.  As can be seen, the overhead introduced by time-slicing in scaled-down mode (right Y-axis) is less than 3\% for most models, demonstrating the efficacy of replica splicing.   Even for small models with low mini-batch times of 200ms, the overhead is only about 5\%.

GPT-2 and InternalT are large models that run with model-parallelism (we used a configuration of 4-way data-parallel, 4-way pipeline-parallel, and 2-way tensor-parallel).  For InternalT, we also used ZERO Stage-1 with our modifications for partial-sharding (2-way sharding of optimizer state).  By efficiently time-slicing on the data-parallel dimension, replica splicing interestingly achieves a small net performance gain over scaled-up execution (because of squashing redundant GPU operations across ranks).  To illustrate the benefit of some of our optimizations, we also performed a run with operation squashing disabled, where, as expected, the overhead of time-slicing was much higher (64\% for BERT, 103\% for GPT-2, 163\% for Int-Q, 72\% for Int-T, 18\% for ResNet/DenseNet).  

Overall, with all our optimizations,  we see that \aisc\ achieves transparent elasticity with negligible overhead, for models across a wide range of sizes and parallelism strategies.

\begin{table}
	\begin{center}
		\footnotesize{
				\begin{tabular}{ l | c| c |  c | c | c| c |c |}
					\hline
					 & \multicolumn{6}{c|}{Latency (s)}  \\
					Model & \multicolumn{2}{c|}{16-to-16} & \multicolumn{2}{c|}{16-to-8} & \multicolumn{2}{c|}{8-to-16}\\
					&  Total & Transfer & Total & Transfer & Total & Transfer \\
					\hline
					BERT & 36 & 17 & 45 & 26 & 34 & 15 \\
					DenseNet169 & 64  & 40 & 69 & 45 &  75 & 51  \\
					PyramidNet & 28  & 7 & 31 & 12 &  30 & 9 \\
					ResNet50 & 58 & 37 & 67 & 45  &  59 & 39  \\
					InternalQ  & 46 & 24 & 68  & 44 & 49 & 24 \\
					GPT-2* & 72 & 44 & 122 & 85  & 97 & 51 \\
					InternalT* & 141 & 98  & 228  & 165 & 181  & 121 \\
					\hline
				\end{tabular}
		}
	\end{center}
	\caption{{\bf Latencies of migration and resizing.} This table shows the end-to-end latency of migration (16-to-16), scale-up (8-to-16) and scale-down (16-to-8).   ($m$-to-$n$) indicates migration from $m$ GPUs to $n$ GPUs. *GPT-2 and InternalT used 32 workers (we report 32-to-32, 32-to-16, and 16-to-32)}
	\vspace{-0.3in}
	\label{table-checkpoint-latency}
\end{table}

\subsection{Latency of migration \& resizing}

Unlike library-based checkpointing where the job is restarted (thus leading to repeated initialization work), migration and resizing in \aisc\ is work-conserving as it resumes from the same program state.  The latency is thus purely a function of the time taken to (a) acquire a barrier, (b) generate a GPU and CRIU dump (c) upload the dump to remote storage, (d) download the dump on a different set of nodes (e) restore the CRIU dumps and GPU dumps, and release the barrier. Table~\ref{table-checkpoint-latency} shows these latencies. For most models, this end-to-end latency for migration or resizing is in tens of seconds, with more than half the time going in the upload and download from remote Azure blob storage (shown as Transfer time).  We are working on a peer-to-peer transfer mechanism over InfiniBand, bypassing the blob store, which would significantly reduce the transfer time in the common case.  Note that these jobs run for hours or even days (\eg, InternalT), so even with several migrations, there is negligible impact on the job runtime.

\subsection{Robustness/Maintainability of device-proxy}

Because of the narrow surface area of our interception and the generality of our domain-specific observations, the device-proxy is robust and maintainable.  In addition to the wide diversity of models explored above, we have tested both the migration and elasticity mechanisms across three versions of PyTorch: 1.6, 1.7, and 1.8, and multiple versions of CUDA: 10.1, 11.0, and 11.1; the effort to get a new version working was negligible, demonstrating the generality and practicality of the approach.  

\section{Related Work}
\label{sec-related}

Prior work related to \aisc\ can be broadly grouped across three categories: DNN schedulers, elasticity mechanisms, and migration mechanisms.

\noindent
{\bf DNN Schedulers. }  Cluster schedulers that are tailored for deep learning have been an actively researched topic in recent times.  Systems such as Gandiva~\cite{Gandiva, GandivaFair}, Tiresias~\cite{tiresias}, Themis~\cite{themis}, Pollux~\cite{pollux} are examples. Most of these systems focus on scheduling policies aimed at optimizing different target metris such as job makespan, JCT, throughput, ``goodput'', \etc.  While some systems~\cite{Gandiva,GandivaFair,hwang2021elastic} propose migration and elastic resource sharing as an enabler for better scheduling, they don't address the problem of how to deal with the vast majority of jobs that are not migratable or elastic.  What differentiates \aisc\ compared to such systems, are the novel and generic mechanisms in \aisc\ to make {\em all} jobs preemptible, resumable and elastic.  These enable \aisc\ to use a new scheduling SLA of GPU fraction that treats preemption and elasticity as first-class constructs.

\noindent
{\bf Job Elasticity. } \aisc\ is the first system that enables efficient, truly transparent elasticity of {\em existing, unmodified} DNN training jobs.  Systems that provide elasticity today for DNN training jobs all require the user to use custom libraries that take over the training loop.  As a result, elasticity is a niche feature today.   \aisc\ represents a paradigm shift by {\em making all jobs elastic}, without any dependency on the job using specific custom libraries.

A recent system~\cite{ElasticityOr} proposes automatic scaling of resources, but it is not transparent, as the effective global batch size of the job varies with the scale, burdening the user with additional complexity.   Systems such as Pytorch-Elastic~\cite{pytorch-elastic} and MXNet Dynamic~\cite{mxnetdl} enable training over variable number of GPUs but they require the user scripts to written in a way that handles resource variability, and to specify different sets of hyper-parameters for each configuration.  A few systems such as Optimus~\cite{optimus}, Gavel~\cite{narayanan2020heterogeneity} and Antman~\cite{antman} automatically determine the best configuration for a job based on maximizing throughput but these systems ignore statistical efficiency  (large number of resources require large batch sizes which can slow down convergence). Pollux~\cite{pollux} is a recent system that models statistical efficiency as well as throughput to determine the best configuration. However, in all these systems, the user still needs to identify the optimal learning rate for their job as a function of batch size.     Instead, with \aisc, elasticity is {\em completely transparent} to the user: the job always runs with the same number of workers with the same parameters; \aisc\ simply remaps these workers to a different number of physical devices.

\noindent
{\bf Job Migration. }  Live process migration has received interest from the systems community for decades~\cite{douglis1991transparent, powell1983process, garg2019mana}.  However, the generality of the workloads that such systems targeted means they were inherently complex, having to deal with the numerous corner-cases that each workload brings, which has made their broad adoption in production settings challenging.  In contrast, migration in \aisc\ is {\em domain-specific}, in that the checkpointing is timed such that the process is in a relatively clean state - \eg, no GPU calls in-flight, no in-flight network communication, aligned to the end of a mini-batch \etc, which makes the migration robust, tractable, and efficient, with negligible steady-state overhead.   By carefully targetting the narrow surface area of CPU-to-GPU communication, \aisc\ employs semantics-aware optimizations (such as delayed error notification).  

In the DNN context, several schedulers employ checkpointing as a primitive, but assume that the application incorporates checkpoint functionality, an unrealistic assumption (\S~\ref{subsec-transparency}). For a service like \aisc\ that uses transparent preemption and elasticity while honoring SLAs, all jobs must be made checkpointable, regardless of whether the user added any code for checkpointing logic explicitly. Thus, low-overhead, transparent, on-demand checkpointing is critical. Gandiva~\cite{Gandiva} was an early system that proposed transparent checkpointing but this was achieved by making extensive changes to the deep learning framework. Given the rapid evolution of these frameworks, such intrusive changes are hard to maintain and keep up to date. In contrast, the proxy-based checkpointing mechanism in \aisc\ is decoupled from the framework, supporting both transparency and maintainability.

\section{Conclusion}
\label{sec-conclusion}

\aisc\ achieves a significant breakthrough in scheduling deep learning workloads, converting niche features such as elasticity into mainstream, always-on features that the scheduler can rely on for implementing stringent SLAs.  With novel mechanisms that make unmodified jobs preemptible and resizable with negligible performance overhead, \aisc\ enables unprecedented levels of workload fungibility, making it possible for jobs to take advantage of spare capacity anywhere in the globally-distributed fleet, while still preserving the SLAs.  \aisc\ achieves all of this with a remarkably simple user experience: the user focuses only on the ML task and does not need to think about checkpointing or elasticity; these mechanisms are infrastructure optimizations that are completely transparent to the user.
\section*{Acknowledgments}

We are grateful to Kevin Scott, Mikhail Parakhin, Peter Lee, and Sriram Rajamani for sharing the vision of \aisc\ and their support. We would like to thank the customers of Singularity for their valuable feedback.
\bibliography{main}
\bibliographystyle{plain}
\end{document}